\documentclass[aps,prl,twocolumn,superscriptaddress,amsmath,amssymb]{revtex4}
\usepackage{graphicx}
\usepackage{amsmath}
\usepackage{amssymb}

\begin{document}

\title{Superradiance of Degenerate Fermi Gases in a Cavity }
\author{Yu Chen}
\author{Zhenhua Yu}
\author{Hui Zhai}
\affiliation{Institute for Advanced Study, Tsinghua University, Beijing, 100084, China}

\date{\today}
\begin{abstract}

In this letter we consider spinless Fermi gases placed inside a cavity and study the critical strength of pumping field for driving a superradiance transition. We emphasize that Fermi surface nesting effect strongly enhances the superradiance tendency. Around certain fillings, when the Fermi surface is nearly nested with a relevant nesting momentum, the susceptibility of the system toward a checkboard density-wave ordered state is strongly enhanced, because of which a much smaller (sometime even vanishingly small) critical pumping field strength can lead to superradiance. This leads to interesting reentrance behavior and topologically distinct structure in the phase diagram. Away from these fillings, the Pauli exclusion principle brings about the dominant effect for which the critical pumping strength is lowered in the low-density regime and increased in the high-density regime, in comparison to a Bose gas with same density. These results open the prospect of studying rich phenomena of degenerate Fermi gases in cavity. 

\end{abstract}
\maketitle

Recently, a series of experiments have studied weakly interacting degenerate Bose gas in a cavity \cite{Esslinger1,Esslinger2}, in which superradiance induced density-ordered superfluid phase \cite{Esslinger1} and softening of roton excitations in the vicinity of a superradiance phase transition have been observed  \cite{Esslinger2}. Studying degenerate quantum gases inside a cavity offers new insights to many-body systems \cite{review}. First, cavity field is a dynamical photon field rather than a classical laser configuration; cavity photon modes affect the many-body system as dynamical variables. For examples, cavity photons can mediate an effective long-range interactions between atoms \cite{long-range,long-range2}; a multi-mode cavity can introduce frustration to atoms that enhances quantum fluctuations \cite{Lev}.  Second, the inevitable decay of cavity photons makes the system interesting for studying 
non-equilibrium phenomena.

So far, limited attention has been paid to degenerate Fermi gases inside cavities \cite{Meystre,Subir,Larson}. However, there is no fundamental difficulty in realizing such a system experimentally. In free space without cavity, superradiance has been proposed theoretically for fermions \cite{Ketterle,Meystre2} and subsequently demonstrated experimentally\cite{Jing}. To stimulate experimental efforts along this direction, it is therefore desirable to theoretically investigate interesting physics in this setup. In this work we shall start from the simplest case, i.e. spinless fermions, and show that nontrivial effects already exist.

In contrast to bosons, due to the Pauli exclusion principle, a degenerate Fermi gas forms a Fermi sea, occupying a collect of single particle states of lowest energies. 
Moreover, the system exhibits a Fermi surface (FS) where ``Fermi surface nesting" is the crucial feature responsible for many collective phenomena in fermonic systems, such as charge-density wave, spin-density wave \cite{nesting}, as well as some strongly correlated unconventional superconductivity \cite{Zhai}. FS nesting means when a sizable portion of the FS shifted by a certain momentum overlaps with the original one. If a FS is perfectly nested, particle-hole excitations of nesting momentum  cost infinitesimally small energies and the FS becomes unstable in the presence of infinitesimally small local repulsive (attractive) interactions and reconstructs to be gapped by spin-density wave (charge-density) wave order.

The purpose of this letter is to point out that FS nesting and the Pauli exclusion principle both have strong effects on superradiance in a degenerate Fermi gas. Explicitly, we show: (i) For the one-dimensional case, perfectly nested FS leads to a dramatic result that infinitesimal pumping field can induce superdiance when the nesting momentum marches the wave-vector magnitude of the cavity field; For the two-dimensional case, close to certain fillings when the nesting momentum marches the momentum transfer $\bf Q$ between the pumping laser and the cavity field photons, superradiance is greatly enhanced. In the nesting regime, the phase diagram exhibits several interesting behaviors.  (ii) In the low density regime, occupation of different single particle states due to the Pauli exclusion principle enhances superradiance; while in the high density regime, superradiance is suppressed by the Pauli exclusion principle.  

\begin{figure}[b]
\includegraphics[width=3.4 in]
{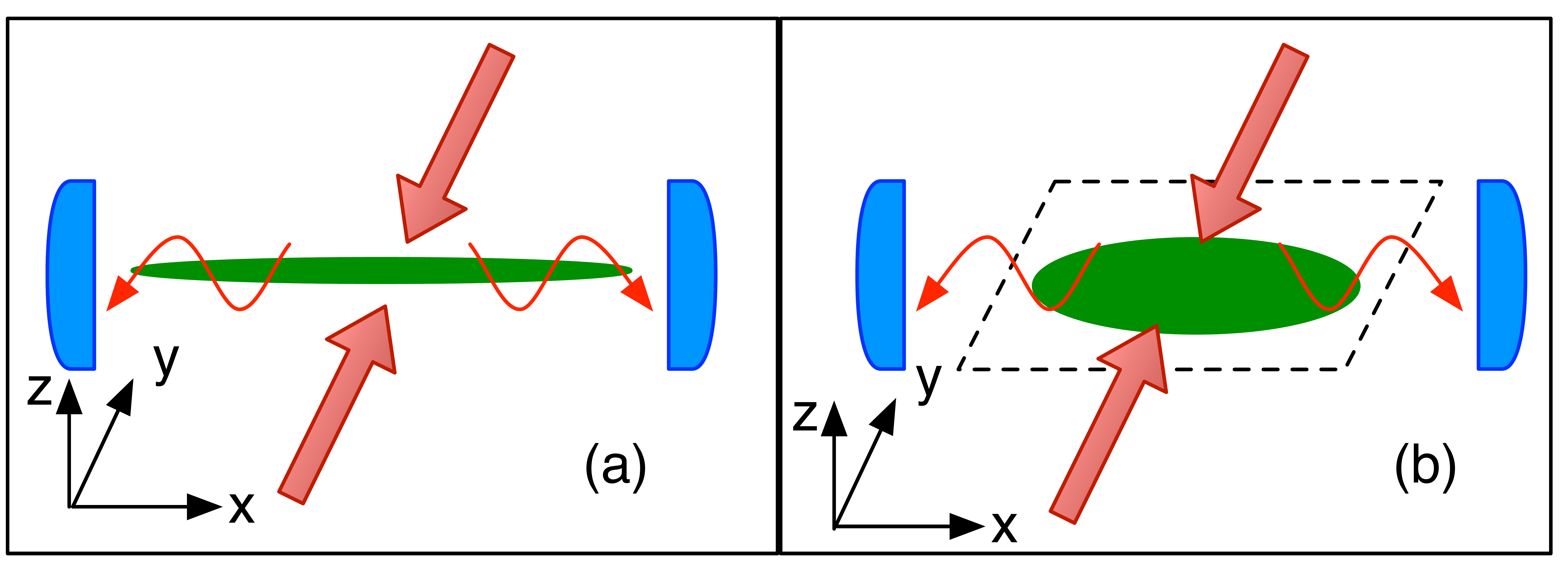}
\caption{Schematic of experimental setup. Arrows are pumping lasers along $\hat{y}$, wiggled lines with arrows represents cavity field along $\hat{x}$. (a) presents a one-dimensional gas along $\hat{x}$, and (b) presents a two-dimensional gas in $xy$ plane with strong confinement along $\hat{z}$. \label{schematic}}
\end{figure}

\textit {Model.} We consider spinless fermionic atoms trapped inside a high-Q cavity. Two linearly polarized pumping laser beams counter-propagate along $\hat{y}$ perpendicular to the main axis $\hat{x}$ of the cavity, as schematically shown in Fig. \ref{schematic}. The gas can be either one-, two- or three-dimensional. For the one-dimensional case, we consider the situation that fermions can only move along the direction of the cavity mode $\hat{x}$, as shown in Fig. \ref{schematic}(a); and for the two-dimensional case, the atoms' motion along $\hat{z}$ is frozen by tight confinement and fermions can only move in the $xy$ plane, as shown in Fig. \ref{schematic}(b). 
The cavity is fine tuned such that only one mode has frequency $\omega_\text{c}$ that is close to the frequency of pumping lasers $\omega_\text{p}$. Both $\omega_\text{p}$ and $\omega_\text{c}$ are far off resonance with respect to the electronic transitions of the atoms so that we can adiabatically eliminate the electronic excited states of the atoms, and obtain the Hamiltonian ($\hbar=1$ throughout) \cite{Esslinger1,model} 
\begin{align}
&\hat{H}=\int d^d{\bf r}\left(\hat{\psi}^{\dag}({\bf r})\hat{H}_0\hat{\psi}({\bf r})\right)-\Delta_c \hat{a}^\dag \hat{a},\label{hami}\\
&\hat{H}_0=\hat{H}_\text{at}+ \eta(\mathbf r)(\hat{a}^\dag+\hat{a})+U(\mathbf r) \hat{a}^\dag \hat{a}\\
&\hat{H}_\text{at}=\frac{{\bf p}^2}{2m}+V({\bf r})
\end{align} 
where $\hat{\psi}$ is the field operators for spinless fermion atoms, and $\hat{a}$ is the field operators for the cavity mode. $V(\mathbf r)$ and $U(\mathbf r)$ are the optical potentials generated by the pumping lasers and the cavity field, respectively, and $V(\mathbf r)=V_0 \cos^2(k_0 y)$, $U(\mathbf r)=U_0 \cos^2(k_0 x)$ with $V_0=\Omega_\text{p}^2/\Delta_\text{a}$ and $U_0=g^2/\Delta_\text{a}$. The interference between the pumping lasers and the cavity field gives rise to $\eta(\mathbf r)=\eta_0\cos (k_0 x)\cos(k_0 y)$ with $\eta_0=g\Omega_\text{p} /\Delta_\text{a}$.
Here $\Delta_\text{a}$ is the laser detuning, $\Delta_\text{c}=\omega_\text{p}-\omega_\text{c}$ is the cavity mode detuning, 
$\Omega_\text{p}$ is the strength of the pumping lasers,
$g$ is the single-photon Rabi frequency of the cavity mode, $k_0$ is the wave-vector magnitude of the pumping lasers and the cavity mode. We define the recoil energy $E_\text{r}=\hbar^2 k^2_0/2m$ for later use. In the following discussion, $g$, $\Delta_\text{a}$, $\Delta_{\rm c}$ and $U_0$ are kept fixed, and superradiance is driven by increasing the pumping field strength $\Omega_{\rm p}$, which simultaneously increases $\eta_0$ via $\eta_0=\sqrt{U_0 V_0}$.

\textit{Method.} The weak leakage of electromagnetic fields from the high-Q cavity leads to a small decay rate $\kappa$ for the cavity mode. The mean field of the cavity mode $\alpha=\langle \hat a\rangle$ satisfies the equation of motion similar as the boson case \cite{Esslinger1}
\begin{eqnarray}
i\frac{\partial \alpha}{\partial t}=\left(-\tilde\Delta_c-i\kappa\right)\alpha+\eta_0 \Theta,
\end{eqnarray}
with the effective detuning $\tilde\Delta_c=\Delta_c-\int d^d{\bf r}U({\bf r})n(\mathbf r)$, the fermion density-order $\Theta=\int d^d{\bf r}n({\bf r})\eta(\mathbf r)/\eta_0$.
The fermion density function is $n(\mathbf r)=\langle\hat\psi^\dagger(\mathbf r)\hat\psi(\mathbf r)\rangle$.
Due to the presence of cavity decay term $\kappa$, the system is generally in a non-equilibrium situation. 
We seek a steady state in which $\partial\alpha/\partial t=0$ and find
\begin{eqnarray}
\alpha=\frac{\eta_0\Theta}{\tilde{\Delta}_c+i\kappa}. \label{steady}
\end{eqnarray}
This steady state requirement fixes the relation between $\alpha$ and $\Theta$.

To determine the critical pumping strength for the superradiance transition, we calculate the free energy by second-order perturbation theory. With the above mean-field treatment for the cavity field and by integrating out the rest fermion fields, we obtain the free energy to the second order of $\alpha$ as
\begin{eqnarray}
F_\alpha=-\beta^{-1}\ln{\cal Z_\alpha}=-\tilde{\Delta}_c\alpha^*\alpha-\chi_\eta(\alpha+\alpha^*)^2, \label{Energy}
\end{eqnarray}
and the susceptibility $\chi_\eta$ is given by 
\begin{equation}
\chi_\eta=-\frac{1}{2\beta}{\rm Tr}[G_0\eta({\bf r}')
G_0\eta({\bf r})]\equiv\eta_0^2 N_\text{at} \frac{f}{E_\text{r}}, \label{sus}
\end{equation}
where $f$ is the dimensionless susceptibility, $N_{\rm at}$ is the total atom number, ${\rm Tr}$ includes the frequency summation and $G_0^{-1}=i\partial_t-H_\text{at}$. Substituting Eq.~ (\ref{steady}) into the free energy expression Eq.~(\ref{Energy}) in the vicinity of the superradiance transition, we obtain
\begin{eqnarray}
F_\alpha=-\left(\frac{\tilde{\Delta}_\text{c}}{\tilde{\Delta}_\text{c}^2+\kappa^2}+\chi_\eta
\frac{4\tilde{\Delta}_\text{c}^2}{(\tilde{\Delta}_\text{c}^2+\kappa^2)^2}\right)(\eta_0\Theta)^2 \label{E2}.
\end{eqnarray}

\begin{figure}[bt]
\includegraphics[width=3.3 in]
{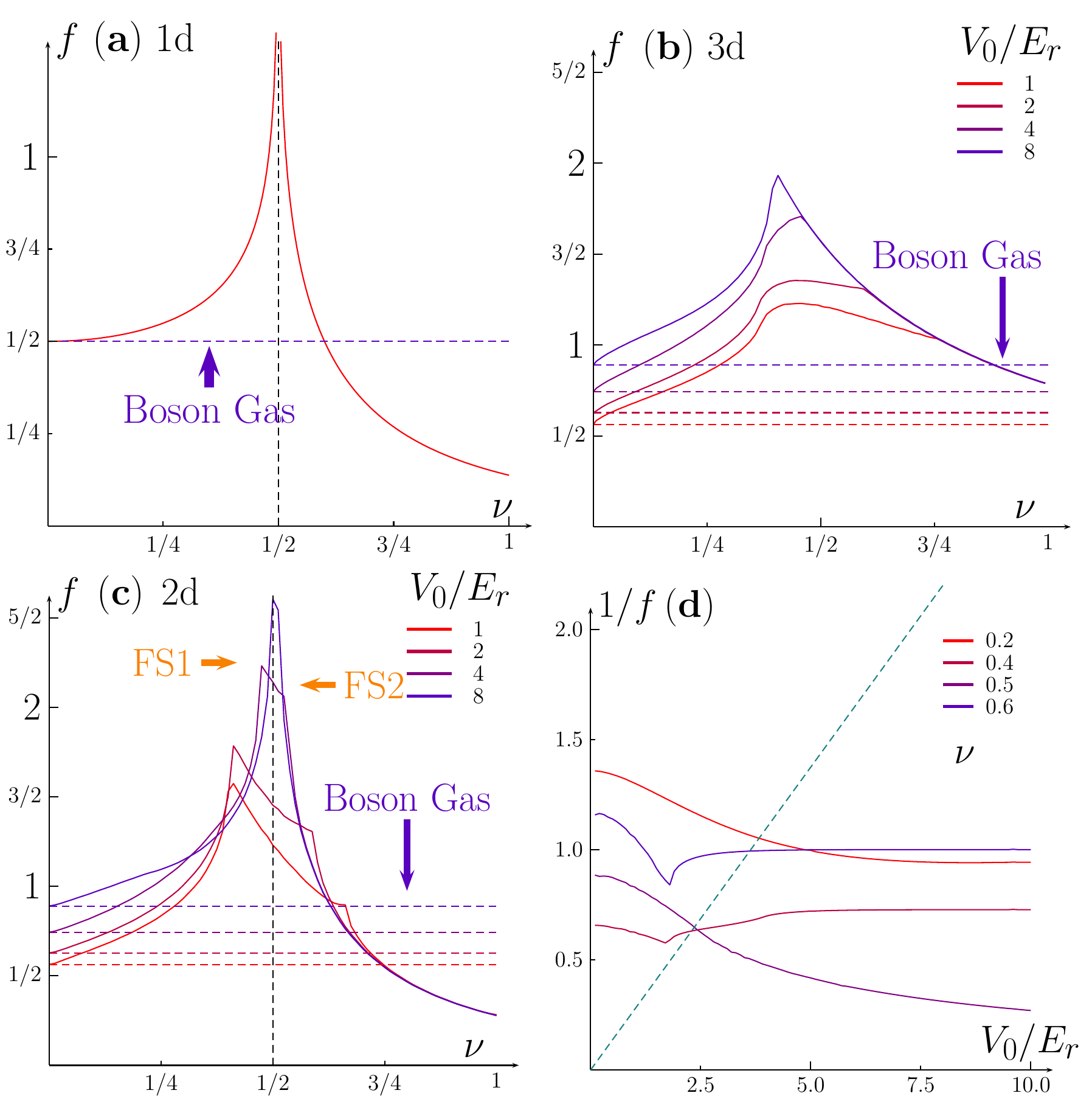}
\caption{(a-c) Dimensionless charge-density-wave susceptibility $f$ is plotted as a function of filling $\nu$, for one-dimension (a); three-dimension (b) and two-dimension (c).  $f$ is defined in Eq. \ref{sus} and is related to the critical pumping strength via Eq. \ref{cri}. In (b) and (c), different lines represent different pumping field strength $V_0/E_\text{r}$. For comparison, horizontal dashed lines represent $f$ for non-interacting boson case with different pumping field strength. (d) $1/f$ as a function of $V_0/E_\text{r}$ for various fillings $\nu$. The dashed line represents $(1/\mathcal{C})V_0/E_\text{r}$ defined in Eq. \ref{critical}. \label{f-plot}}
\end{figure}

Across a superradiance transition, $\Theta$ simultaneously evolves from zero to a finite value. Therefore the transition is determined by the sign change of the coefficient of $\Theta^2$ in Eq.~(\ref{E2}), which yields the critical value of $\eta_0$
\begin{equation}
\eta^{\rm cr}_0\sqrt{N_\text{at}}=\frac{1}{2}\sqrt{\frac{\tilde{\Delta}_\text{c}^2+\kappa^2}{-\tilde{\Delta}_\text{c}}}\sqrt{\frac{E_\text{r}}{f}}.\label{cri}
\end{equation}
It is straightforward to show that in terms of the eigen-functions $\phi_{\bf k}$ and the eigen-energies $\epsilon_{\bf k}$ of $\hat{H}_\text{at}$,
\begin{equation}
f =\frac{E_\text{r}}{\eta^2_0N_{\rm at}}\sum_{{\bf k}{\bf k^\prime}}\left|\int d^d\mathbf r \phi^*_{\bf k}(\mathbf r)\phi_{\bf k^\prime}(\mathbf r)\eta({\bf r})\right|^2\frac{n_\text{F}(\epsilon_{\bf k})}{\epsilon_{\bf k^\prime}-\epsilon_{\bf k}}\label{f-function},
\end{equation}
with $n_\text{F}$ the Fermi distribution function. For fermion case, $f$ depends on the atom density, the pumping field strength and the dimensionality of the atomic gas. For boson case, the critical value is also determined by Eq.~(\ref{cri}), and the difference is that in the expression for $f$, $n_{\rm F}(\epsilon_{\bf k})$ in Eq.~(\ref{f-function}) should be replaced by $N_{\rm at}\delta_{{\bf k},0}$ at zero-temperature, where ${\bf k}=0$ corresponds to the Bose condensed single particle ground state. Thus, for non-interacting boson case, $f$ is independent of atom density and $f\approx 1/2$ for weak pumping field \cite{Esslinger1}. The magnitude of $f$ determines the easiness of inducing superradiance. The larger $f$, the smaller critical pumping strength. We present the numerical results for $f$ at zero temperature based on Eq.~(\ref{f-function}) in Fig. \ref{f-plot} for Fermi gases at different dimensions and compare them with non-interacting Bose gases.

\textit {Low Density and High Density Limit.} If the filling factor $\nu\equiv n /(\alpha k^d_0)$, with $\alpha=2$(for $d=1$) and $\alpha=4$ (for $d=2,3$)
 for a $d$-dimensional Fermi gas of average density $n$, is much smaller than one, the degenerate fermions mainly occupy the lowest lying single particle states. At zero temperature, in the limit $\nu\to0$, one finds that $f$ approaches the same value for bosons and fermions \cite{Ketterle, Meystre2,note}. And this values increases as the lattice depth increases, which means lattice effect enhances the superradiance tendency, as shown in Fig. \ref{f-plot}(b) and (c). In fact, similar effect has also been found in resonance physics where lattice effect enhances the tendency of molecule formation \cite{Zoller,Cui}. When $\nu$ increases from zero, $f$ for fermions becomes larger than that for bosons, while the later remains unchanged due to its independence of atoms' density.
The increment of $f$ for fermions comes from the population of finite momentum states, since some of the finite momentum states have a smaller energy denominator in the Eq.~(\ref{f-function}).

We also find that as $\nu$ increases to the high density regime with $2k_\text{F}>|{\bf Q}|$ (${\bf Q}=(\pm k_0,\pm k_0)$ for two-dimension case and ${\bf Q}=(\pm k_0,\pm k_0,0)$ for three-dimension case), $f$ for fermions will finally drop below that for bosons. This is because when the Fermi surface is large enough, for a certain number of occupied states with momentum ${\bf k}$, the states with momentum ${\bf k}+{\bf Q}$ will also be occupied and the Pauli exclusion principle blocks the scattering between these states. The superradiance tendency is suppressed accordingly.

\textit{Nesting Effect.} The nesting effect can be best illustrated in the one-dimensional case, where $f$ can be calculated analytically as (up to a constant \cite{c})
\begin{align}
f=&\frac{1}{8}\frac{k_0}{k_\text{F}}\ln\left|\frac{2k_\text{F}+k_0}{2k_\text{F}-k_0}\right|,
\end{align}
at zero-temperature. As shown in Fig. 2(a), $f$ diverges when $k_0=2k_\text{F}$, which means that an infinitesimally small pumping field can lead to superradiance. The divergence is due to that in one-dimension, FS contains only two points and is generically nested with the nesting momentum $2k_\text{F}$. The interaction mediated by cavity photons $\sim\cos(k_0x)$ can only transfer a fixed momentum $k_0$. Thus, only when $2k_\text{F}$ marches $k_0$, infinitesimal cavity-mediated attraction between fermions is capable of inducing a density-wave order of fermions. Finite temperature is expected to smear out the divergence of $f$ and results in a finite critical strength for the pumping field. 

\begin{figure}[bt]
\includegraphics[width=3.4 in]
{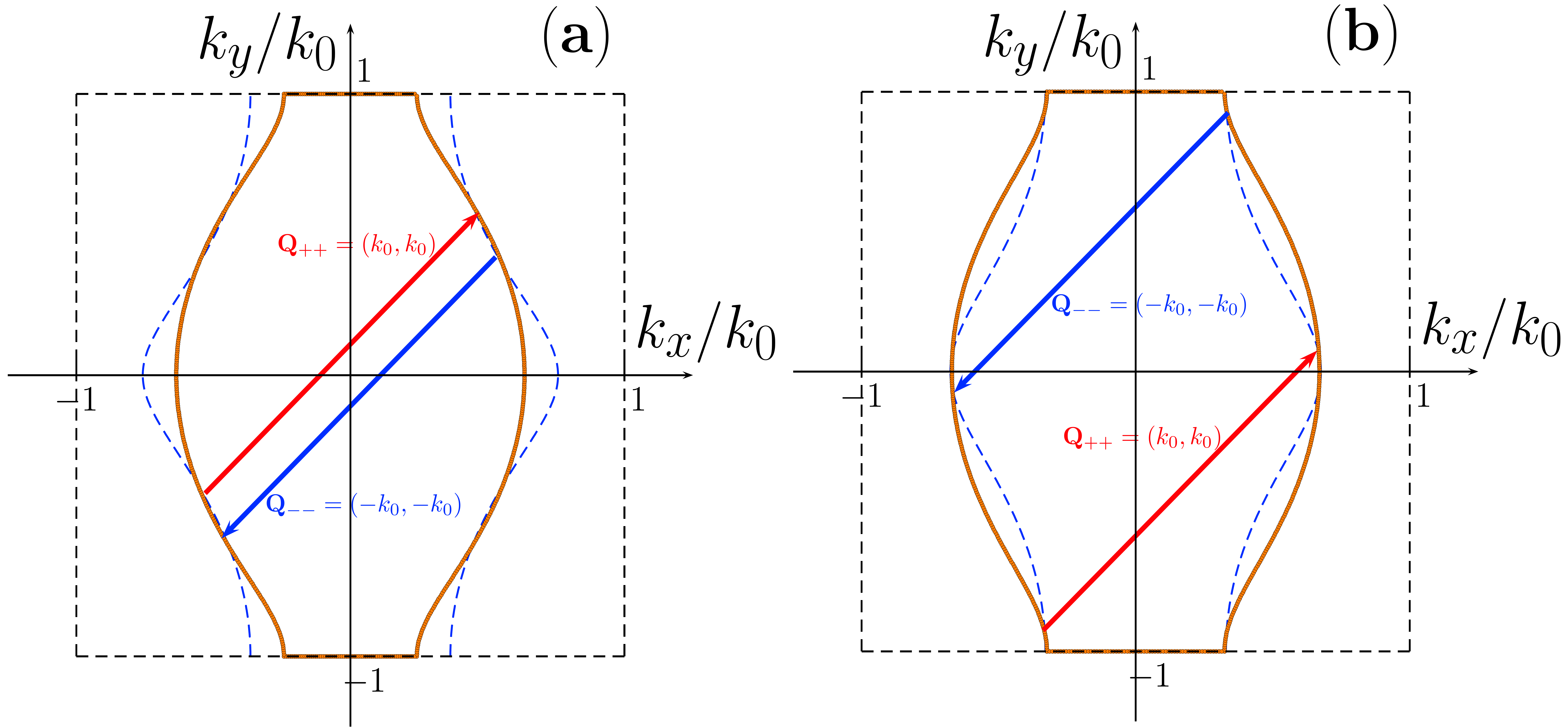}
\caption{Nesting of FS for Hamiltonian $\hat{H}_\text{at}$ before superradiance takes place, with two different fillings as marked in Fig. \ref{f-plot}(c). Solid line is the original FS, and dashed lines are FS shifted by ${\bf Q}=(\pm k_0,\pm k_0)$. Arrows indicate the momentum transfer $\bf Q$. \label{FS}}
\end{figure}

While in higher dimension, FS is generally not nested, and the higher dimension, the more difficult to find a nested FS. In two-dimension, there are still cases that a sizable portion of FS is nearly nested. When the nesting momentum roughly marches ${\bf Q}$, $f$ will be largely increased although remains finite. For the two-dimension case, as shown in Fig. 2(c), $f$ as a function of $\nu$ displays two peaks in the regime $\nu\approx 0.5$ and the exact locations of these peaks depend on pumping field strength. In Fig. \ref{FS}, we plot the FS for $\hat{H}_\text{at}$ at these peak positions. Indeed we find that part of the FS is well nested with relevant momentum ${\bf Q}$, which proves that the nested FS is responsible for the significant increasing of $f$ in the two-dimension case. Similarly, a peak around $\nu\approx 1/2$ is found in three-dimension case, as shown in Fig. \ref{f-plot}(b).

\textit{Determining Phase Diagram.} For bosons or fermions of a given density in one-dimension, $f$ is independent of $V_0/E_\text{r}$. The phase boundary $V^\text{cr}_0/E_\text{r}$ as a function of $\tilde{\Delta}_\text{c}/E_\text{r}$ can be derived directly from Eq.~(\ref{cri}) if $\kappa/E_\text{r}$ and $\sqrt{U_0 N_\text{at}/E_\text{r}}$ are given. For fermions in higher dimensions, $f$ is also a function of $V_0/E_\text{r}$. To determine the phase boundary one needs to solve the equation
\begin{align}
\frac{V_0^\text{cr}}{E_\text{r}}=\frac{\mathcal C(\tilde{\Delta}_{\rm c}/E_\text{r})}{f\left(V^\text{cr}_0/E_\text{r}\right)} \label{critical}
\end{align}
where
\begin{equation}
\mathcal C(x)=\frac{1}{4}\left(\frac{x^2+(\kappa/E_\text{r})^2}{-x}\right)\left(\frac{1}{U_0N_\text{at}/E_\text{r}}\right).
\end{equation}
In Fig. \ref{f-plot}(d), we plot $1/f$ as a function of $V_0/E_\text{r}$ and a straight line representing $V_0/E_\text{r}\mathcal C(\tilde{\Delta}_{\rm c}/E_\text{r})$, whose crossing marks the superradiance transition point. 

\begin{figure}[bt]
\includegraphics[width=3.4 in]
{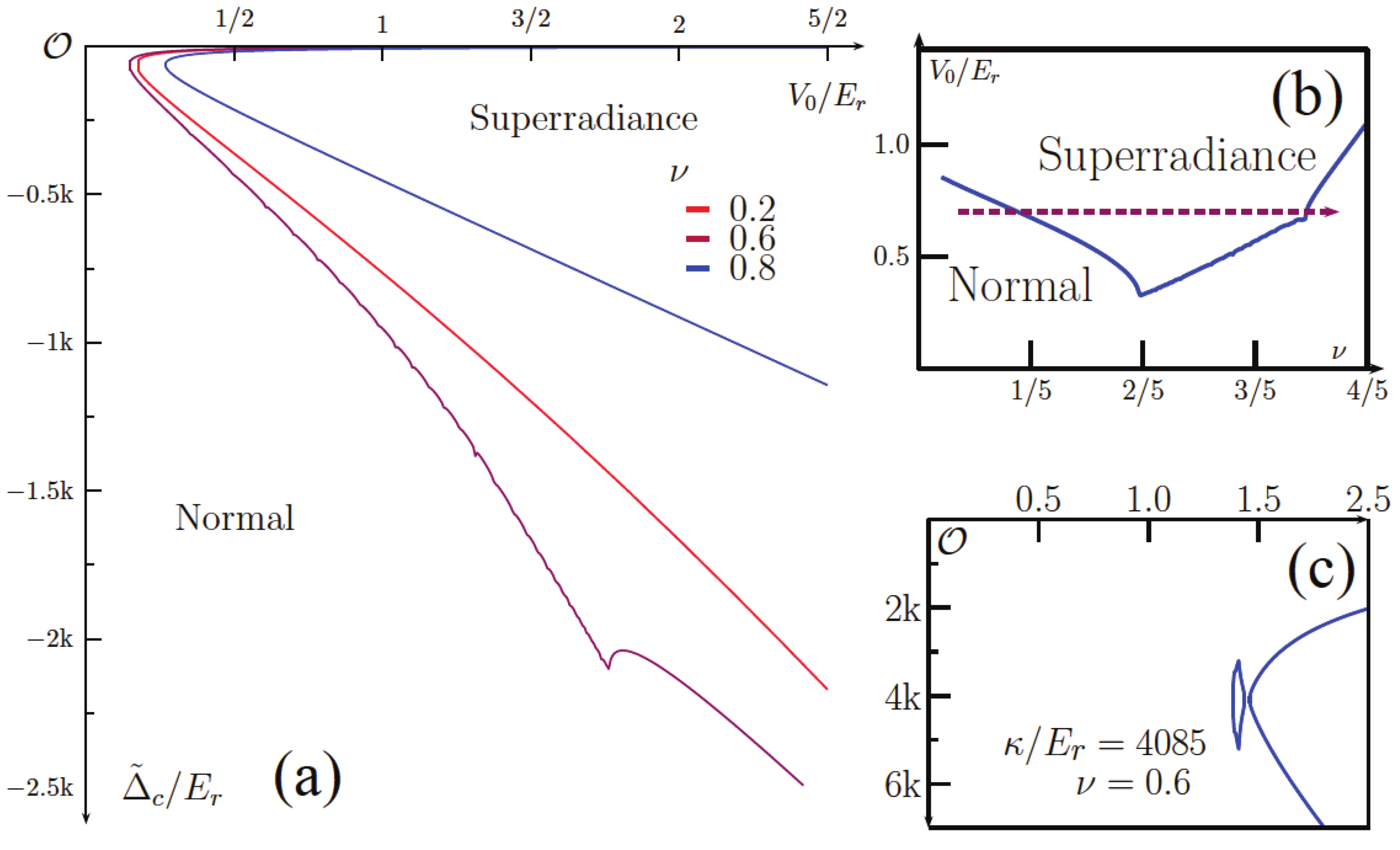}
\caption{(a) and (c): The phase diagram for two-dimension case, in terms of effective detuning $\tilde{\Delta}_\text{c}/E_\text{r}$ and pumping lattice depth $V_0/E_\text{r}$. Different lines in (a) represent phase boundary with different fillings. (b) Critical $V_0/E_\text{r}$ as a function of filling $\nu$ for $\tilde{\Delta}/E_\text{r}$ fixed at $2\times 10^3$. $\kappa/E_\text{r}=250$ for (a) and (b);   $\kappa/E_\text{r}=4085$ for (c) and $U_0N_\text{at}/E_\text{r}=1\times 10^3$ for (a-c). \label{Phase-Diagram}}
\end{figure}

In Fig. \ref{Phase-Diagram} we plot the phase diagram for different densities; the curves are the boundary separating the normal and the superradiance phases. For a fixed effective detuning $\tilde{\Delta}_\text{c}/E_\text{r}$, the critical pumping strength $V_0/E_\text{r}$ is shown to reach its minimum
in the nesting regime $\nu\approx 1/2$.  In another word, there is density-driven superradiance transition and reentrance behavior as shown in Fig. \ref{Phase-Diagram}(b): the system starting in the normal phase undergoes a transition to the superradiance phase and comes back to the normal phase as the density further increases.
In addition, due to the non-monotonic behavior of $1/f$ for filling $\nu\approx 1/2$, for certain fine-tuned $\kappa/E_\text{r}$, the phase diagram can exhibit topologically distinct behavior as shown in Fig. \ref{Phase-Diagram}(c), where an additional isolated island of superradiance regime exists in the phase diagram.

\textit{Final Remark.} In this work we have revealed that many-body effects have much stronger impact on the superradiance of degenerate Fermi gases in a cavity, even for spinless fermions and single mode cavity. Though the quantitative results we have shown are for zero temperature, the enhancement of superradiance for fermions compared to bosons is expected to maintain at finite temperatures. Our results lay the base for further efforts to understand more intriguing phenomena in this system, for instance, by including the fluctuations of cavity modes, considering multiple cavity modes or interactions between fermions of different spin-degree of freedom.

{\it Acknowledgements.} This work is supported by Tsinghua University Initiative Scientific Research Program, NSFC under Grant No. 11004118, No. 11174176, No. 11104157 and No. 11204152, and NKBRSFC under Grant No. 2011CB921500.

{\it Note Added.} During completing this work, we became aware of two papers by J. Keeling, M. J. Bhaseen, and B. D. Simons, arXiv:1309.2464 and F. Piazza and P. Strack, arXiv:1309.2714, in which similar issue is addressed.


\begin{thebibliography}{99}
\bibitem{c}
The constant is $\int dydz\varphi^2_0(y,z)\cos(k_0 y)$ where $\varphi_0(y,z)$ is the wave-function for the lowest mode in the confinement direction.
\end{thebibliography}
\end{document}